# High Q-factor CMOS-MEMS inductor

Ching-Liang Dai, Jin-Yu Hong and Mao-Chen Liu
Department of Mechanical Engineering,
National Chung Hsing University,
Taichung, 402 Taiwan, R.O.C.

*Abstract*-This study investigates a high Q-factor spiral inductor fabricated by the CMOS (complementary metal oxide semiconductor) process and a post-process. The spiral inductor is manufactured on silicon substrate using the 0.35 μm CMOS process. In order to reduce the substrate loss and enhance the Q-factor of the inductor, silicon substrate under the inductor is removed using a post-process. The post-process uses RIE (reactive ion etching) to etch the sacrificial layer of silicon dioxide, and then TMAH (tetra methyl ammonium hydroxide) is employed to remove the underlying silicon substrate and obtain the suspended spiral inductor. The advantage of the post process is compatible with the CMOS process. The Agilent 8510C network analyzer and a Cascade probe station are used to measure the performances of the spiral inductor. Experiments indicate that the spiral inductor has a Q-factor of 15 at 11 GHz, an inductance of 4 nH at 25.5 GHz and a self-resonance frequency of about 27 GHz.

## I. INTRODUCTION

Micro inductors that are important components can be applied in LC tank, VCO [1] and DC-DC converters [2]. The Q-factor is an important characteristic for inductors. The energy dissipation in inductors depends on the Q-factor of inductors. As the Q-factor of inductors increases, the energy dissipation in inductors decreases. Several studies have used MEMS (microelectromechanical system) technology to enhance the Q-factor of micro inductors. For instance, the electroplated solenoid-type inductors, presented by Nam et al. [3], were fabricated on both a standard silicon substrate and glass substrate by thick PR photolithography and copper electroplating. The maximum Q-factor of the inductors was about 10, and the inductance varied range from 1nH to 5nH. Tseng et al. [4] employed the post-CMOS process to fabricate micro inductors. The post-CMOS process adopted an anisotropic $CF_4/O_2$ RIE dry etching to remove the dielectric oxide layer, and then an isotropic $SF_6/O_2$ RIE dry etching was utilized to etch the underlying silicon substrate and release the suspended inductor. The maximum Q-factor of the inductors was 15, and the inductance was 1.88 nH at 8.5 GHz. Dai et al. [5] manufactured a micro suspended inductor using CMOS-MEMS process. The suspended inductor was released by a post-process after it completed the CMOS process. The post-process used etchants to etch the sacrificial layers of metal, and then TMAH was applied to remove the underlying silicon substrate and release the suspended inductor. The maximum Q-factor of the inductor was 4.7. An edge-suspended inductor, proposed by Chen et al. [6], was fabricated using a combination of deep dry etching and anisotropic wet etching techniques. The inductor had an inductance of 4.5-nH, a maximum Q-factor of 11.7 and a self-resonance frequency of 14.3 GHz. Fang et al. [7] developed a three-dimensional micro inductor with air-core using the surface micromachining and electroplating techniques. The maximum Q-factor of the inductor was 22.9 and the inductance was 1.17 nH at 5.5 GHz. Ahn et al. [8] used the surface micromachining technique to make a solenoid inductor with electroplated nickel-iron permalloy cores on silicon wafer. The Q-factor and inductance of the solenoid inductor were 1.5 at 1 MHz and 0.1 μH at 10 kHz, respectively.

The technique that utilizes the commercial CMOS process to fabricate MEMS devices is known as CMOS-MEMS [9-11]. The benefits of micro devices fabricated by the CMOS-MEMS technique are compatible with the CMOS process and easy mass-production. In this work, we employ the CMOS-MEMS technique to manufacture a spiral inductor. In order to enhance the Q-factor of the inductor, a post-process is adopted to remove silicon substrate under the inductor. The post-process utilizes $CHF_3/O_2$ RIE to etch the sacrificial layer of silicon dioxide, and then TMAH is used to remove the underlying silicon substrate. Experiments show that the suspended spiral inductor has a Q-factor of 15 at 11 GHz and an inductance of 4 nH at 25.5 GHz.

## II. DESIGH AND FABRICATION

Figure 1 shows a planar spiral inductor, where $D$ is the internal diameter of the spiral inductor, $W$ is the wire width of the spiral inductor and $S$ is the spacing between the wires of the spiral inductor. In this investigation, the planar spiral inductor is designed as $D$=136 μm, $W$=10 μm, $S$= 2 μm, and the number of turns is 3.5.

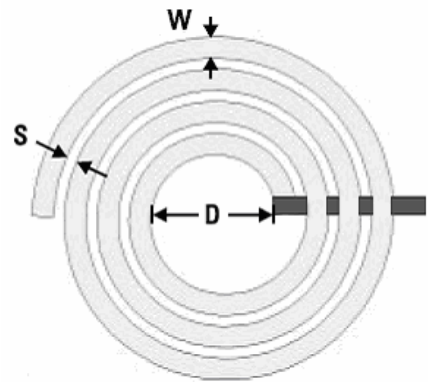

Fig. 1 Structure of the spiral inductor.

The Q-factor of the inductors, which measures the capability of the inductors to save energy, is an important parameter. According to the $\pi$ model as shown in Fig. 2, the Q-factor of the spiral inductors is given by [12],

$$Q = \frac{\omega L_s}{R_s} \cdot \frac{R_p}{R_p + \left[ \left( \omega L_s / R_s \right)^2 + 1 \right] R_s}$$

$$\cdot \left[ 1 - \frac{R_s^2 (C_s + C_p)}{L_s} - \omega^2 L_s (C_s + C_p) \right]$$

$$= \frac{\omega L_s}{R_s} \cdot \left( \begin{array}{c} \text{Substrate loss} \\ \text{factor} \end{array} \right) \cdot \left( \begin{array}{c} \text{Self-resonance} \\ \text{factor} \end{array} \right) \quad (1)$$

where

$$R_p = \frac{1}{\omega^2 C_{ox}^2 R_{Si}} + \frac{R_{Si}(C_{ox} + C_{Si})^2}{C_{ox}^2} \quad (2)$$

$$C_p = C_{ox} \cdot \frac{1 + \omega^2 (C_{ox} + C_{Si}) C_{Si} R_{Si}^2}{1 + \omega^2 (C_{ox} + C_{Si})^2 R_{Si}^2} \quad (3)$$

and $\omega$ represents frequency. The second term in Eq. (1) is the substrate loss factor representing the energy dissipated in the silicon substrate. Therefore, the Q-factor of the inductor depends on the substrate loss. The substrate loss of the inductors can be improved in two ways. The first increases the distance between the spiral inductor and silicon substrate surface [13]. The second method is to increase the resistivity of the silicon substrate to prevent the loss of current in the silicon substrate [14]. In this work, we utilize the first method to reduce the substrate loss. A post-CMOS process is utilized to etch the underlying silicon substrate, and to increase the distance between the spiral inductor and silicon substrate surface.

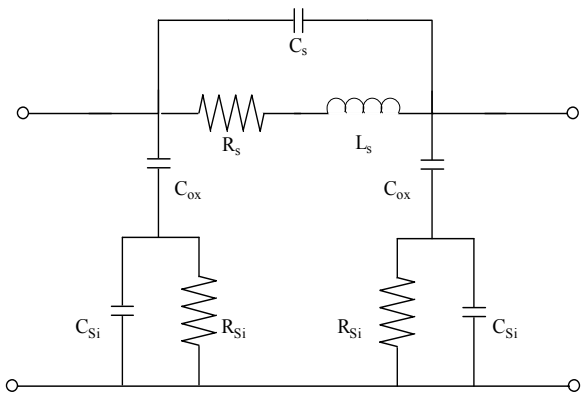

Fig. 2 $\pi$ model for the spiral inductor.

The spiral inductor is fabricated on silicon substrate using the CMOS process of TSMC (Taiwan Semiconductor Manufactur

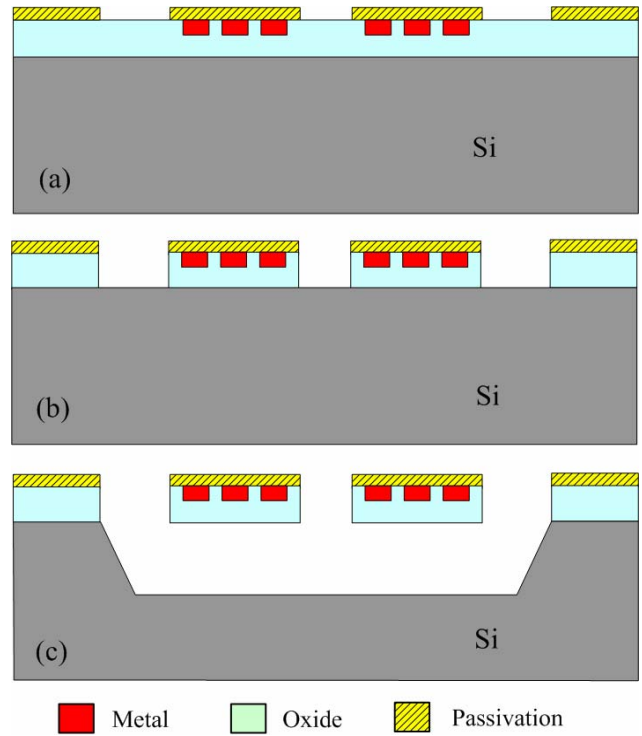

Metal    Oxide    Passivation

Fig. 3 Fabrication flow of the spiral inductor; (a) after the COMS process, (b) oxide layer etched by an anisotropic dry etching, (c) silicon substrate etched by an anisotropic wet etching.

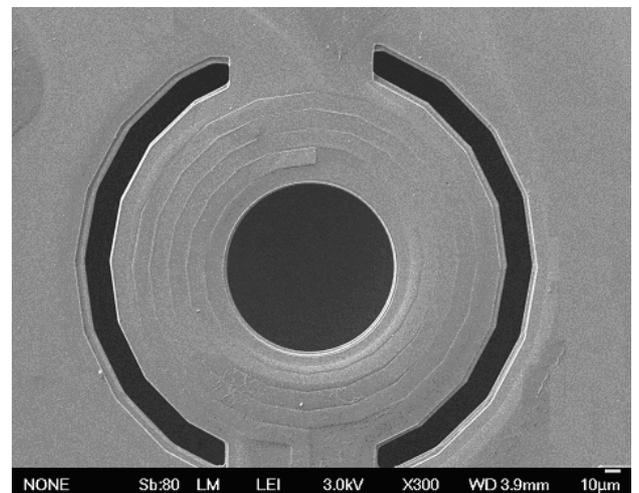

Fig. 4 SEM image of the suspended spiral inductor after the post-process.

-ing Company). Figure 3 illustrates the fabrication flow of the spiral inductor. Figure 3(a) shows the spiral inductor after completion of the CMOS process. The material of the spiral inductor is metal layer. In order to reduce the substrate loss and enhance the Q-factor of the inductor, the post-process is adopted to remove the underlying silicon substrate. The post-process consists of two steps. One is to etch the sacrificial layer, and the other step is to remove silicon substrate. Figure 3(b) presents the sacrificial layer etched by a dry etching. The sacrificial layer is silicon dioxide. The anisotropic $CHF_3/O_2$ RIE dry etching is used to etch the sacrificial layer, and to expose silicon substrate. Figure 3(c) shows the underling silicon substrate etched by a wet etching. The etchant of TMAH at the temperature of 70 °C is applied to remove the underlying silicon substrate, and to form the suspended spiral inductor. Figure 4 depicts the SEM (scanning electron microscope) image of the spiral inductor after completion of the post-process. The post-process is compatible with the CMOS process.

## III.   RESULTS AND DISCUSSION

The Agilent 8510C network analyzer and a Cascade probe station were used to measure the characteristics of the spiral inductor. In order to characterize the performances of the spiral inductor in high operating frequency, the test pad of the spiral inductor adopts ground-signal-ground (GSG) structure to connect the Cascade probe station. Hence, the parasitic effect of pad needs to remove by using the de-embedding procedure. A dummy open pad was designed to remove the parasitic effect.

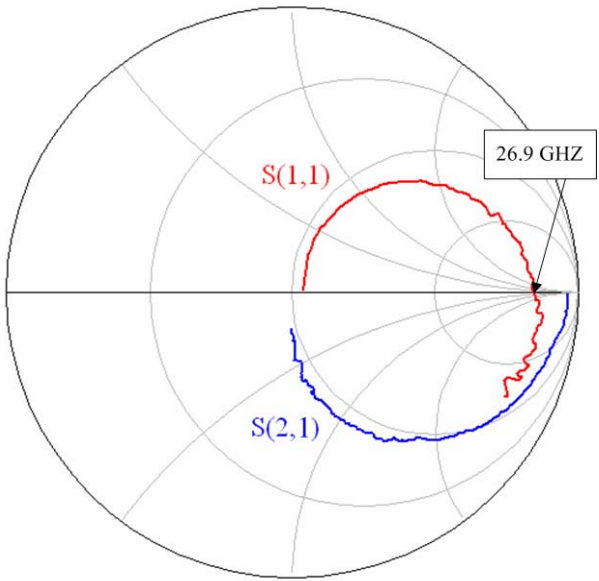

Fig. 5 Smith chart of the spiral inductor.

The spiral inductor was measured in the range of 0.1-40 GHz. Figure 5 shows the Smith chart of the spiral inductor, where S11 and S21 represent the S-parameters of the inductor. The S-parameters at microwave frequency are the reflection and transmission coefficients. The reflection coefficient (S11) is directly related to impedance, and the transmission coefficient (S21) is commonly called gain or attenuation. As shown in Fig. 5, the curve of S11 parameter located in the upper semicircle of the Smith chart, so the device displayed an inductance characterization. The spiral inductor had a self-resonance frequency of about 27 GHz. Figure 6 presents the Q-factor of the spiral inductor after de-embedding procedure. The results revealed that the maximum Q-factor of the inductor was 15 at 11 GHz. Figure 7 illustrates the inductance of the spiral inductor after de-embedding procedure. The measurement depicted that the inductance of the inductor varied range from 1.2 nH to 4 nH at 0.1-25.5 GHz. Thereby, the experimental results showed that the spiral inductor had a Q-factor of 15 at 11 GHz, an inductance of 4 nH at 25.5 GHz and a self-resonance frequency of 27 GHz.

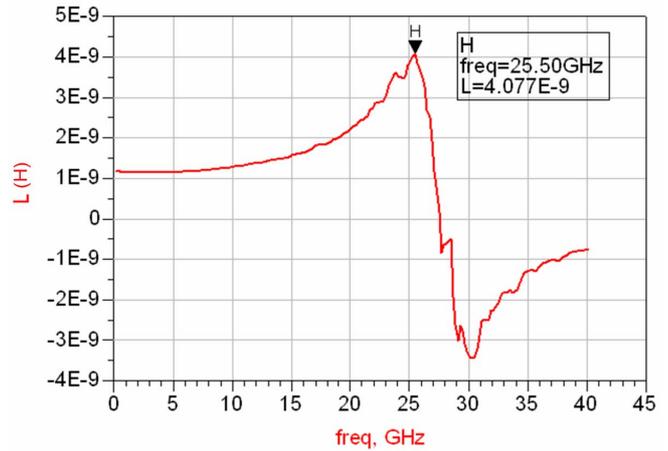

Fig. 6 Inductance of the spiral inductor.

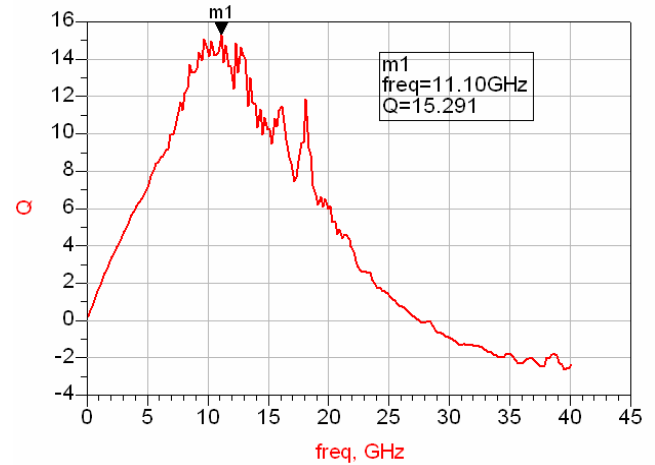

Fig. 7 Q-factor of the spiral inductor.

Park et al. [14] adopted high-resistivity silicon substrate to improve the Q-factor of micro inductors. The spiral inductors with the rectangular and circular shape were built on the 2 kΩ·cm silicon substrate using the conventional COMS process without any post-process, and the maximum Q-factor of the inductors was 12. Lakdawala et al. [15] used the 0.18 μm 6-level copper CMOS process and a post-CMOS process to fabricate a suspended spiral inductor. The post-CMOS process applied an anisotropic $CHF_3/O_2$ RIE dry etching and an isotropic $SF_6/O_2$ RIE dry etching to etch the dielectric layer and silicon substrate, respectively, to release the suspended spiral inductor, in which the maximum Q-factor of the spiral inductor was 7. With the same post-process, Tseng et al. [4] manufactured the suspended inductors applied in a 5.8 GHz VCO, in which the maximum Q-factor of the suspended inductors was 15. Dai et al. [5] employed the commercial 0.35 μm CMOS process and a post-process to produce a suspended inductor. The post-process adopted etchants to remove the sacrificial layers of metal, and then TMAH was utilized to etch the underlying silicon substrate and release the suspended inductor, in which the maximum Q-factor of the suspended inductor was 4.7. Ozgur et al. [16] made a spiral inductor on the suspended membrane using the 1.2 μm CMOS process and a post-process. In the post-process, an isotropic silicon etchant (XeF2) was used to etch the backside of silicon substrate, and to form the suspended membrane. The spiral inductor located on the suspended membrane, and the maximum Q-factor of the inductor was 10.5. In this work, the maximum Q-factor of the inductor was 15. A comparison of the above literatures, the maximum Q-factor of this work is the same as that of Tseng et al. [4] and exceeds that of Dai et al. [5], Park et al. [14], Lakdawala et al. [15] and Ozgur et al. [16].

## IV. CONCLUSION

The high Q-factor suspended spiral inductor fabricated using the 0.35 μm CMOS process and a post-process has been implemented. In this work, we utilized the post-process to remove the underlying silicon substrate, and to reduce the substrate loss and enhance the Q-factor of the inductor. The post-process included two steps. One was to use $CHF_3/O_2$ RIE to etch the sacrificial layer of silicon dioxide, and to expose silicon substrate. The other step was to apply TMAH at the temperature of 70 °C to etch the underlying silicon substrate, and to form the suspended spiral inductor. Experimental results showed that the inductance of the suspended spiral inductor varied range from 1.2 nH to 4 nH at 0.1-25.5 GHz, and the self-resonance frequency and maximum Q-factor of the spiral inductor were 27 GHz and 15, respectively. The maximum Q-factor of this work exceeded that of Dai et al. [5], Park et al. [14], Lakdawala et al. [15] and Ozgur et al. [16]. The benefit of the post-process was compatible with the CMOS process. Thereby, the suspended inductor was capability of integrating with radio-frequency (RF) integrated circuits on a chip.

## ACKNOWLEDGMENT

The authors would like to thank National Center for High-performance Computing (NCHC) for chip simulation, National Chip Implementation Center (CIC) for chip fabrication and the National Science Council of the Republic of China for financially supporting this research under Contract No NSC 95-2221-E-005-043-MY2.